\documentclass[letterpaper]{article} 
\usepackage[preprint]{aaai2027}
\usepackage[hyphens]{url} 
\usepackage{graphicx} 
\usepackage{natbib} 
\usepackage{caption} 
\usepackage{amsmath}
\usepackage{amssymb}
\usepackage{booktabs}
\usepackage{multirow}
\newcommand{\Description}[1]{}

\title{RenderFormer++: Scalable and Physics-Informed Feed-Forward Neural Rendering}
\author{
Huangsheng Du\equalcontrib, Haoran Zhu\equalcontrib, Youcheng Cai\corresponding\\
Jingyang Meng, Ligang Liu
}
\affiliations{
University of Science and Technology of China\\
Hefei, China\\
\texttt{caiyoucheng@ustc.edu.cn}
}

\begin{document}
\maketitle

\begin{abstract}

   We present RenderFormer++, a scalable and physics-informed feed-forward neural rendering framework for global illumination in mesh scenes. Existing Transformer-based neural rendering methods such as RenderFormer achieve promising cross-scene generalization, but lack explicit transport priors and scale poorly due to quadratic triangle-level attention. To address these issues, we introduce Physics-Informed Transport Guidance (PITG), which embeds rendering-equation-inspired inductive biases into the attention mechanism and introduces a transport consistency loss, encouraging physics-informed light transport modeling. We further propose Hierarchical Object-Centric Tokenization (HOCT), which aggregates triangle-level features into compact object-level tokens via cross-attention with learnable queries, substantially reducing computational and memory costs. Extensive experiments demonstrate that RenderFormer++ achieves scalable and generalizable feed-forward global illumination rendering across complex large-scale scenes with competitive rendering quality and substantially improved efficiency over RenderFormer. The code will be made publicly available upon acceptance.

\end{abstract}

\section{Introduction}
\label{sec:intro}
Modeling global illumination for scenes represented as discrete triangle meshes remains a fundamental challenge in computer graphics. Classical rendering methods rely on physically based simulations of light transport and can accurately reproduce complex phenomena such as indirect illumination, soft shadows, and specular interreflections. More recently, neural rendering \cite{tewari2022Advances} has emerged as a promising data-driven paradigm for modeling global illumination. In particular, transformer-based approaches, such as RenderFormer \cite{zeng2025renderformer}, offer the possibility of feed-forward rendering models that generalize across diverse mesh scenes without requiring per-scene optimization.

\begin{figure}[t]
  \centering
  \includegraphics[width=0.95\linewidth]{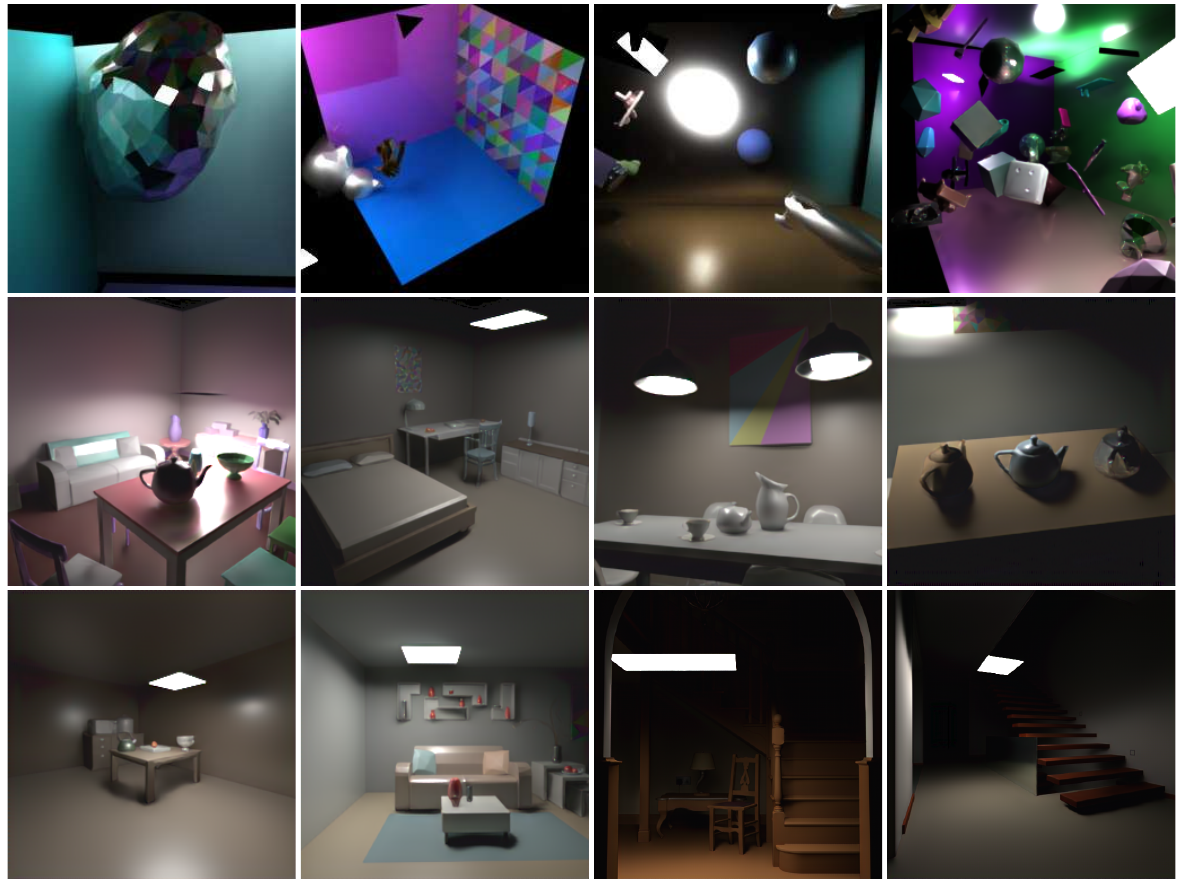}
  \caption{Feed-forward global illumination results produced by RenderFormer++ across diverse complex scenes with varied geometry, materials, and lighting conditions.}
  \label{fig:teaser}
\end{figure}

Existing neural rendering methods for global illumination can generally be categorized into two paradigms. Neural radiance methods \cite{hadadan2021neural, muller2021Realtime} model global illumination through continuous or learned scene-specific spatial functions. Neural transport methods \cite{diolatzis2022Active, ren2024lightformer} embed physical constraints into their training objectives or architectures to better simulate complex light transport. Despite achieving high rendering quality, methods in both paradigms are typically trained or optimized in a scene-specific manner. This inherent dependence on per-scene training severely limits their ability to generalize to unseen and diverse environments.

To overcome this generalization barrier, recent feed-forward architectures such as RenderFormer \cite{zeng2025renderformer} represent an important step toward fully neural and generalizable rendering pipelines for triangle-mesh scenes. By tokenizing scene triangles into attention-based representations, RenderFormer captures global light transport interactions with minimal prior assumptions, allowing the network to learn transport behaviors directly from data. However, this purely data-driven paradigm suffers from two major limitations: (1) \textit{Physical Constraints}: relying solely on attention mechanisms does not explicitly enforce the physical laws governing light transport as described by the rendering equation \cite{kajiya1986rendering}; and (2) \textit{Scalability}: naive triangle-level tokenization causes computational and memory costs to grow quadratically with the number of scene triangles. This poor scalability restricts applicability to large-scale scenes.

In this work, we seek to address these limitations and enable high-fidelity neural rendering for substantially larger and more complex environments. The primary challenges are twofold: (1) how to effectively incorporate the inductive biases of the rendering equation into Transformer-based light transport modeling; and (2) how to design a more efficient tokenization strategy that replaces naive triangle-based tokenization while substantially reducing computational overhead and preserving essential geometric and radiometric information.

To address these challenges, we propose \textbf{RenderFormer++}, a scalable and physics-informed feed-forward neural rendering framework for global illumination. First, we introduce \textbf{Physics-Informed Transport Guidance (PITG)}, which explicitly incorporates rendering-equation-inspired inductive biases into the Transformer architecture. Specifically, we embed light transport principles directly into the attention mechanism and introduce a transport consistency loss to minimize a rendering-equation-inspired residual in feature space during training, encouraging physics-informed light transport modeling. Second, we propose a \textbf{Hierarchical Object-Centric Tokenization (HOCT)} strategy that aggregates triangle-level tokens into compact object-level tokens through cross-attention with learnable query vectors. This abstraction significantly reduces the sequence length processed by the attention layers, effectively alleviating the quadratic computational bottleneck and enabling scalable rendering for large multi-object scenes. Finally, during the view-dependent decoding stage, we introduce a geometry-guided decoder that integrates G-buffers with aggregated object-level tokens to produce high-fidelity renderings. Through extensive experiments on complex mesh scenes with diverse geometric structures and lighting configurations, we demonstrate that the proposed framework enables scalable and generalizable feed-forward global illumination modeling.

Our primary contributions are summarized as follows:
\begin{itemize}
\item We propose \textbf{RenderFormer++}, a scalable, physics-informed, and feed-forward neural rendering framework capable of generalizing across complex, large-scale triangle-mesh environments.
\item We introduce two key technical innovations: \textbf{PITG}, which incorporates rendering-equation-inspired inductive biases for physics-informed light transport modeling, and \textbf{HOCT}, which alleviates the quadratic computational bottleneck through compact object-centric tokenization, jointly enabling efficient and physics-informed rendering of complex environments.
\end{itemize}

\section{Related Work}

\subsection{Physically-Based Light Transport}
\label{Physically-Based Light Transport}


Physically based rendering models image formation through radiance transport. The rendering equation \cite{kajiya1986rendering} expresses outgoing surface radiance as the sum of emitted radiance and reflected incident radiance integrated over the hemisphere. Solving it requires accounting for recursive interactions among surfaces and light sources, modulated by reflectance functions such as BRDFs \cite{nicodemus1977geometrical}, which produce global illumination effects.


Monte Carlo path tracing is the standard approach to approximating the rendering equation, but its stochastic estimation suffers from high variance and slow convergence. Neural importance sampling \cite{dong2023Neural, huang2024Online} and radiance caching \cite{muller2021Realtime, majercik2019Dynamic} improve sampling and reuse efficiency, yet remain within the conventional Monte Carlo pipeline.


Radiosity methods \citep{sillion1991Global, goralComputer, immelRadiosity} provide a finite-element formulation of light transport but are largely limited to diffuse surfaces. Neural Radiosity and its dynamic-scene extensions demonstrate the value of physically grounded inductive biases \citep{hadadan2021neural, su2024Dynamic, coomans2024Realtime}. However, these methods do not address unified feed-forward rendering for complex triangle-mesh scenes, motivating our rendering-equation-inspired design.

\subsection{Neural Rendering}

Recent advances in neural rendering have explored replacing traditional rendering pipelines with neural network-based inference of light transport. Existing approaches can be broadly grouped into three categories. The first category focuses on scene-specific neural approximations of global illumination, including Neural Radiosity and its dynamic extensions \citep{hadadan2021neural, su2024Dynamic, coomans2024Realtime}. These methods produce high-quality results but require per-scene fitting and therefore offer limited cross-scene generalization. The second category models the light transport process through neural scene representations \citep{granskog2020Compositional, eslami2018Neural}. Specifically, active exploration improves coverage of variable scenes \citep{diolatzis2022Active}, while regression-based methods approximate scene-specific transport and extend it to dynamic area lights \citep{ren2013global, gao2023Neural}. Furthermore, source-to-object and inter-object transport interactions are explicitly modeled in \citep{ren2024lightformer, zheng2024Neural}. Despite their differences, these approaches remain largely scene-specific. The third category represents a departure from the above methods. RenderFormer tokenizes mesh triangles and models their interactions with self-attention, enabling cross-scene feed-forward rendering \citep{zeng2025renderformer}. However, it remains constrained by light and camera distributions and the number of scene primitives, leaving scalable rendering for complex meshes an open challenge.

\subsection{Transformer-based feed-forward network}

Transformer architectures \cite{vaswaniAttention} use attention to model long-range dependencies and have proven effective in vision \cite{dosovitskiy2021IMAGE}, language modeling \cite{devlinBERT}, and 3D shape representation \cite{zhang20233DShape2VecSet}. More recently, models such as VGGT \cite{wangVGGT} and AnySplat \cite{jiang2025AnySplat} have demonstrated their potential for feed-forward structured inference by amortizing multi-stage geometric inference, post-optimization, or per-scene reconstruction into a single forward pass that predicts scene geometry, camera parameters, or view-synthesis-ready representations. These advances position transformers as learnable surrogates for structured inference rather than merely generic feature extractors. In rendering, attention-based architectures have similarly been used to model light transport \cite{ren2024lightformer}. Most closely related, RenderFormer \cite{zeng2025renderformer} formulates triangle-mesh rendering as a feed-forward sequence-to-sequence transformation, using self-attention to capture global transport interactions across scene primitives. Although it enables cross-scene generalization, its flat triangle-level tokenization becomes a scalability bottleneck as scene complexity increases.

\section{Method}


We consider the problem of rendering a high-dynamic-range (HDR) image with global illumination directly from an explicit triangle-mesh representation of a scene. The 3D scene is represented as a collection of $M$ triangles, denoted as $\mathcal{S}=\{t^m\}_{m=1}^{M}$. Each triangle $t^m$ is associated with geometric and radiometric attributes, including spatial coordinates, shading normals, surface reflectance properties (e.g., diffuse albedo, specular albedo, and roughness), as well as emission profiles. Our objective is to learn a feed-forward neural mapping function $f_{\theta}$ that takes the explicit scene representation $\mathcal{S}$ and the camera pose $\mathcal{C}$ as inputs and directly predicts the target HDR image $\mathbf{I} \in \mathbb{R}^{3 \times H \times W}$:
\begin{equation}
\mathbf{I} = f_{\theta}(\mathcal{S}, \mathcal{C}).
\end{equation}

RenderFormer \cite{zeng2025renderformer} adopts a similar formulation by representing $\mathcal{S}$ as triangle-level tokens, learning transport interactions directly over scene primitives, and decoding image values conditioned on $\mathcal{C}$. In this work, we investigate how rendering-equation-inspired inductive biases can be incorporated into Transformer-based light transport modeling, while simultaneously designing a more efficient scene encoding strategy to alleviate the computational bottleneck in complex scenes.

\begin{figure*}[t]
  \centering
  \includegraphics[width=\textwidth]{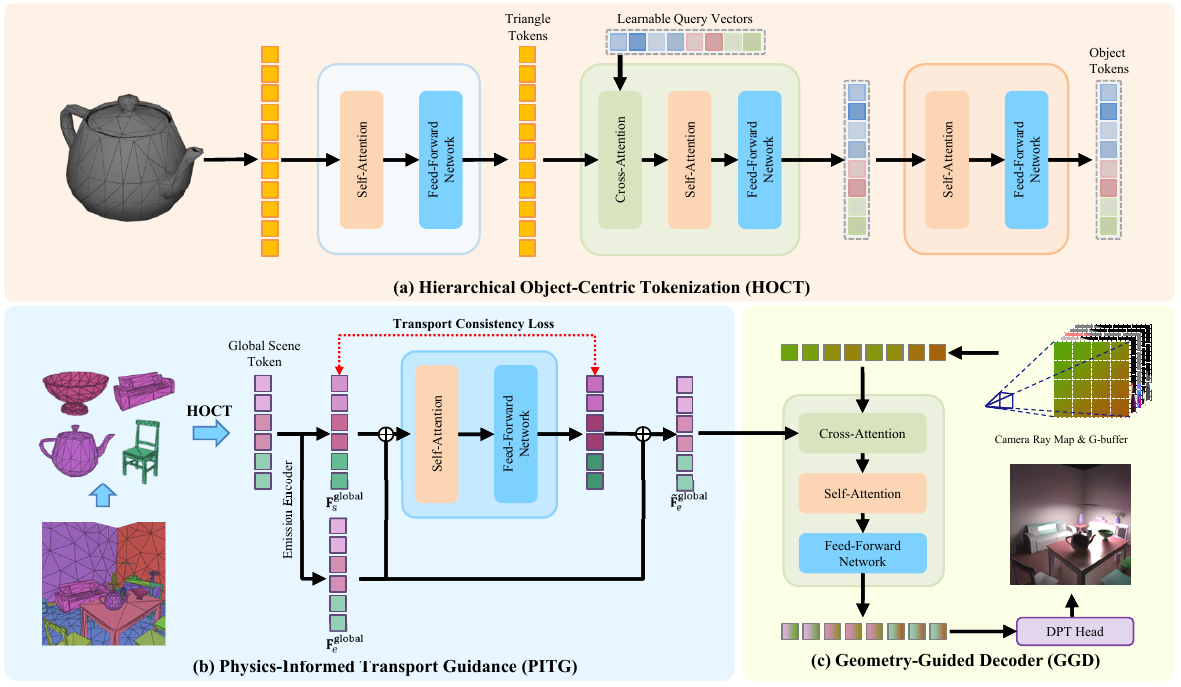}
  \caption{Overview of RenderFormer++. The framework first aggregates triangle-level features into compact object-level tokens, refines them through physics-informed transport modeling, and decodes view-dependent radiance using geometry-guided queries.}
  \Description{Pipeline diagram of RenderFormer++, including hierarchical object-centric tokenization, physics-informed transport guidance, and geometry-guided radiance decoding.}
  \label{fig:pipeline}
\end{figure*}

Figure~\ref{fig:pipeline} provides an overview of RenderFormer++, which first aggregates per-object triangle features into compact object-level tokens using HOCT, refines the resulting global representation with PITG, and then predicts the HDR image through the geometry-guided decoder.

\subsection{Physics-Informed Transport Guidance}

PITG takes as input a global scene token $\mathbf{F}^{\text{global}}$. Our goal is to model light transport directly through the attention mechanism of Transformers. In physically based rendering, the light transport operator, denoted as $T$, propagates radiance through the scene by transporting incident radiance to the next visible surface, where it is scattered according to the surface reflectance. Following \citet{soler2022theoretical}, the equilibrium radiance field satisfies the rendering equation:
\begin{equation}
\mathbf{L}_o = T\mathbf{L}_o + \mathbf{L}_e,
\label{lte}
\end{equation}
where $\mathbf{L}_o$ denotes the outgoing radiance field, $\mathbf{L}_e$ represents emitted radiance from light sources, and $\mathbf{L}_s=T\mathbf{L}_o$ corresponds to transported and scattered radiance within the scene.

RenderFormer~\cite{zeng2025renderformer} implicitly models Eq.~\ref{lte} as:
\begin{equation}
\mathbf{L}_o = T_\theta(\mathbf{F}^{\text{global}}),
\end{equation}
where $T_\theta$ denotes the Transformer-based neural renderer parameterized by $\theta$. In RenderFormer, the transport process is learned in a largely unconstrained manner from triangle-based scene tokenization and global attention interactions.

However, such a direct formulation lacks explicit physical structure for representing key properties of light transport, especially the separation between emitted and transported radiance and the recursive nature of multi-bounce illumination. To address this limitation, we introduce Physics-Informed Transport Guidance (PITG), which injects rendering-equation-inspired inductive biases into the Transformer architecture through three complementary mechanisms: (1) an emission encoder that explicitly incorporates emission priors into feature representations, (2) a neural light transport operator that reflects the recursive structure of light transport in feature space, and (3) a transport consistency loss that regularizes iterative transport dynamics during training.

\paragraph{\textbf{Emission Encoder.}}
In the rendering equation, emissive surfaces act as the origins of illumination and provide the source term for global light transport. To mirror this behavior within our neural formulation, we explicitly inject emission priors into the global scene token $\mathbf{F}^{\text{global}}$.

Let $\mathbf{E}$ denote the emission features aligned with the scene tokens, where entries corresponding to non-emissive elements are zero-padded. We concatenate $\mathbf{E}$ with the scene tokens and map the resulting features into the transport feature space via an emission encoder, denoted as $\mathrm{EEncode}(\cdot)$ and implemented using multi-layer perceptrons (MLPs). This transformation produces the global emission token $\mathbf{F}_e^{\text{global}}$ and the global non-emission token $\mathbf{F}_s^{\text{global}}$:
\begin{equation}
\begin{aligned}
\mathbf{F}_e^{\text{global}} &= \mathrm{EEncode}\big(\mathrm{concat}(\mathbf{F}^{\text{global}}, \mathbf{E})\big), \\
\mathbf{F}_s^{\text{global}} &= \mathrm{EEncode}\big(\mathrm{concat}(\mathbf{F}^{\text{global}}, \mathbf{0})\big),
\end{aligned}
\end{equation}
where $\mathbf{0}$ denotes an all-zero emission embedding.

This explicit emission injection is applied once before the neural light transport stage, ensuring that localized light source information is embedded into the global token representation. Intuitively, $\mathbf{F}_e^{\text{global}}$ represents features associated with emitted radiance $\mathbf{L}_e$, while $\mathbf{F}_s^{\text{global}}$ represents features corresponding to transported and scattered radiance $\mathbf{L}_s$.

\paragraph{\textbf{Neural Light Transport Operator.}}
Inspired by Neural Radiosity~\cite{hadadan2021neural}, which formulates global illumination as an iterative transport process, we rewrite Eq.~\ref{lte} as:
\begin{equation}
\mathbf{L}_s = T(\mathbf{L}_e + \mathbf{L}_s).
\end{equation}
This formulation emphasizes that scattered radiance is recursively generated from the combination of emitted radiance and previously transported radiance. To preserve this operator-based structure in feature space, we design a Transformer-based neural light transport operator $T_\theta$:
\begin{equation}
\mathbf{F}_s^{\text{global}}
= T_\theta\left(
\mathbf{F}_s^{\text{global}} + \mathbf{F}_e^{\text{global}}
\right).
\end{equation}

Here, the self-attention mechanism serves as a learnable feature-space transport mechanism that propagates illumination information across scene tokens, enabling long-range interactions for global illumination prediction in a fully feed-forward manner.

Finally, the refined scattered features are combined with the emission features to obtain the final global illumination representation:
\begin{equation}
\tilde{\mathbf{F}}^{\text{global}}
=
\mathbf{F}_s^{\text{global}}
+
\mathbf{F}_e^{\text{global}}.
\end{equation}

Compared with unconstrained Transformer-based rendering of RenderFormer, our formulation introduces physics-informed structure into feature propagation, separately encodes features with and without explicit emission conditioning, and supports the modeling of complex illumination effects.

To model recursive light transport in feature space, we repeatedly apply the neural light transport operator using shared parameters across iterations. At iteration $k$, the transport update is defined as:
\begin{equation}
{}^{(k+1)}\mathbf{F}_s^{\text{global}}
=
T_\theta\left(
{}^{(k)}\mathbf{F}_s^{\text{global}}
+
\mathbf{F}_e^{\text{global}}
\right),
\end{equation}
where the same operator $T_\theta$ is shared across all iterations.

This recursive refinement progressively propagates illumination information throughout the scene, allowing the network to iteratively refine global illumination representations in feature space while maintaining parameter efficiency. In our implementation, we empirically set the number of transport iterations to $K=3$. The influence of the iteration $K$ on rendering quality is further analyzed in the ablation studies.

\paragraph{\textbf{Transport Consistency Loss.}}
Motivated by the recursive structure of the rendering equation, we introduce a transport consistency loss to regularize intermediate transport states during training.

Specifically, we define the objective as the cumulative discrepancy between the current transport state and the updated state predicted by the neural transport operator:
\begin{equation}
\mathcal{L}_{\text{cons}}
=
\sum_{k=1}^{K}
\left\|
{}^{(k)}\mathbf{F}_s^{\text{global}}
-
T_\theta\big(
{}^{(k)}\mathbf{F}_s^{\text{global}}
+
\mathbf{F}_e^{\text{global}}
\big)
\right\|_2^2.
\end{equation}


This regularization penalizes discrepancies between transport states and their predicted updates, encouraging consistent multi-step feature refinement during training while preserving a feed-forward inference mechanism at test time.

\subsection{Hierarchical Object-Centric Tokenization}

Directly modeling light transport over unstructured sequences of triangles yields a sequence length that scales linearly with the total number of primitives in a scene. This formulation inherently limits scalability as geometric complexity increases. To address this computational bottleneck while preserving essential structural information, we introduce a Hierarchical Object-Centric Tokenization (HOCT) strategy.

\paragraph{\textbf{Triangle Embedding.}} 
Triangle-mesh scenes are naturally organized into objects, denoted as $\mathcal{S}=\{O_n\}_{n=1}^{N}$, where each object $O_n$ contains a subset of scene triangles. For a given mesh object $O_n = \{t_n^m\}_{m=1}^{M_n}$, we initially treat its constituent triangles as an unordered set. Following the design of RenderFormer \cite{zeng2025renderformer}, the spatial coordinates of each triangle are encoded using Relative Spatial Positional Encoding (RSPE). Concurrently, we encode intrinsic geometric and radiometric attributes, including shading normals, emission profiles, and surface reflectance parameters. Surface reflectance is modeled using a microfacet BRDF with a GGX normal distribution function (NDF) \cite{walterMicrofacet}, parameterized by diffuse albedo, specular albedo, and roughness values.

The combination of positional and attribute embeddings yields a sequence of encoded triangle tokens for object $O_n$, denoted as $\mathbf{H}_n$. This primitive-level encoder shares its weights across all objects to ensure consistent extraction of local geometric structures.

\paragraph{\textbf{Object-Centric Tokenization.}} 

To reduce the sequence length processed by the global transport module, we aggregate the triangle tokens of each object via cross-attention. Inspired by 3DShape2VecSet~\cite{zhang20233DShape2VecSet}, we introduce a learnable query set $\mathbf{Q} \in \mathbb{R}^{n_q \times d}$ shared across all
objects, where $n_q$ denotes the number of learnable queries and determines the number of output object-level tokens per object, and $d$ denotes the feature dimension. The shared queries provide a common aggregation basis for summarizing geometric and radiometric information across objects.

We use $\mathbf{Q}$ as the query embeddings in a cross-attention operation over the encoded triangle tokens $\mathbf{H}_n$. For a given object $O_n$, the token aggregation process is formulated as:
\begin{equation}
\mathbf{F}_n = \mathrm{CrossAttn}(\mathbf{Q}, \mathbf{H}_n),
\end{equation}
where $\mathbf{F}_n$ represents the compact fixed-length representation of object $O_n$. Finally, the global scene token is constructed by concatenating all object-level token sequences: $ \mathbf{F}^{\text{global}} = \mathrm{Concat}(\{\mathbf{F}_n\}_{n=1}^{N})$.

\subsection{Geometry-Guided Decoder}

To predict the final view-dependent radiance, we adopt the patch-based decoding architecture of RenderFormer \cite{zeng2025renderformer}. The target image is partitioned into non-overlapping $4 \times 4$ patches. For each patch $p$, we construct a query token by encoding a bundle of primary rays that pass through the centers of its constituent pixels.

To provide local scene cues for decoding, we extract the corresponding G-buffer attributes (e.g., surface normals, albedo, and roughness) from a pre-rendered G-buffer image, as commonly adopted in prior rendering methods \cite{ren2024lightformer, zheng2024Neural}. The ray vectors and their spatially aligned G-buffer attributes are concatenated and jointly embedded to form a geometry-guided patch token $\mathbf{V}$. We design a Transformer-based decoder, denoted as $\mathrm{VDecoder}$, to synthesize the HDR radiance for each patch by conditioning on both the local geometry-guided token $\mathbf{V}$ and the globally aggregated light transport tokens $\tilde{\mathbf{F}}^{\text{global}}$:

\begin{equation}
\mathbf{I} = \mathrm{VDecoder}(\mathbf{V}, \tilde{\mathbf{F}}^{\text{global}}),
\end{equation}
where $\mathbf{I}$ represents the predicted HDR image. By incorporating G-buffer attributes into the ray-bundle queries, the decoder fuses local geometric and material details with the global illumination resolved by the transport operator.

\subsection{Network Architecture}


RenderFormer++ comprises view-independent and view-dependent
stages. In the view-independent stage, a four-layer self-attention Transformer first encodes the triangle sequence of each object. HOCT then uses $n_q=8$ learnable queries to aggregate these triangle tokens through three cross-attention layers and two self-attention layers, producing compact object-level tokens. The object-level tokens are concatenated into a global scene sequence. An emission encoder then generates global emission and non-emission tokens, which are fed into a two-layer neural transport operator. In the view-dependent stage, a six-layer geometry-guided Transformer decoder and a DPT head predict the HDR image. RenderFormer++ has a parameter count comparable to that of RenderFormer.

\paragraph{\textbf{Loss Function}} We train the network end-to-end in a supervised manner. Our final optimization objective combines an $L_1$ loss in log-transformed HDR space between the rendered reference image and our predicted image, an LPIPS loss to minimize perceptual discrepancies, and the transport consistency loss:
\begin{equation}\mathcal{L} = \mathcal{L}_{1} + \lambda_{\text{lpips}} \mathcal{L}_{\text{lpips}} + \lambda_{\text{cons}} \mathcal{L}_{\text{cons}}
\end{equation}
where we empirically set the weighting hyperparameters to $\lambda_{\text{lpips}} = 0.05$ and $\lambda_{\text{cons}} = 0.1$.

\begin{figure*}[t]
  \centering
  \includegraphics[width=0.95\textwidth,keepaspectratio]{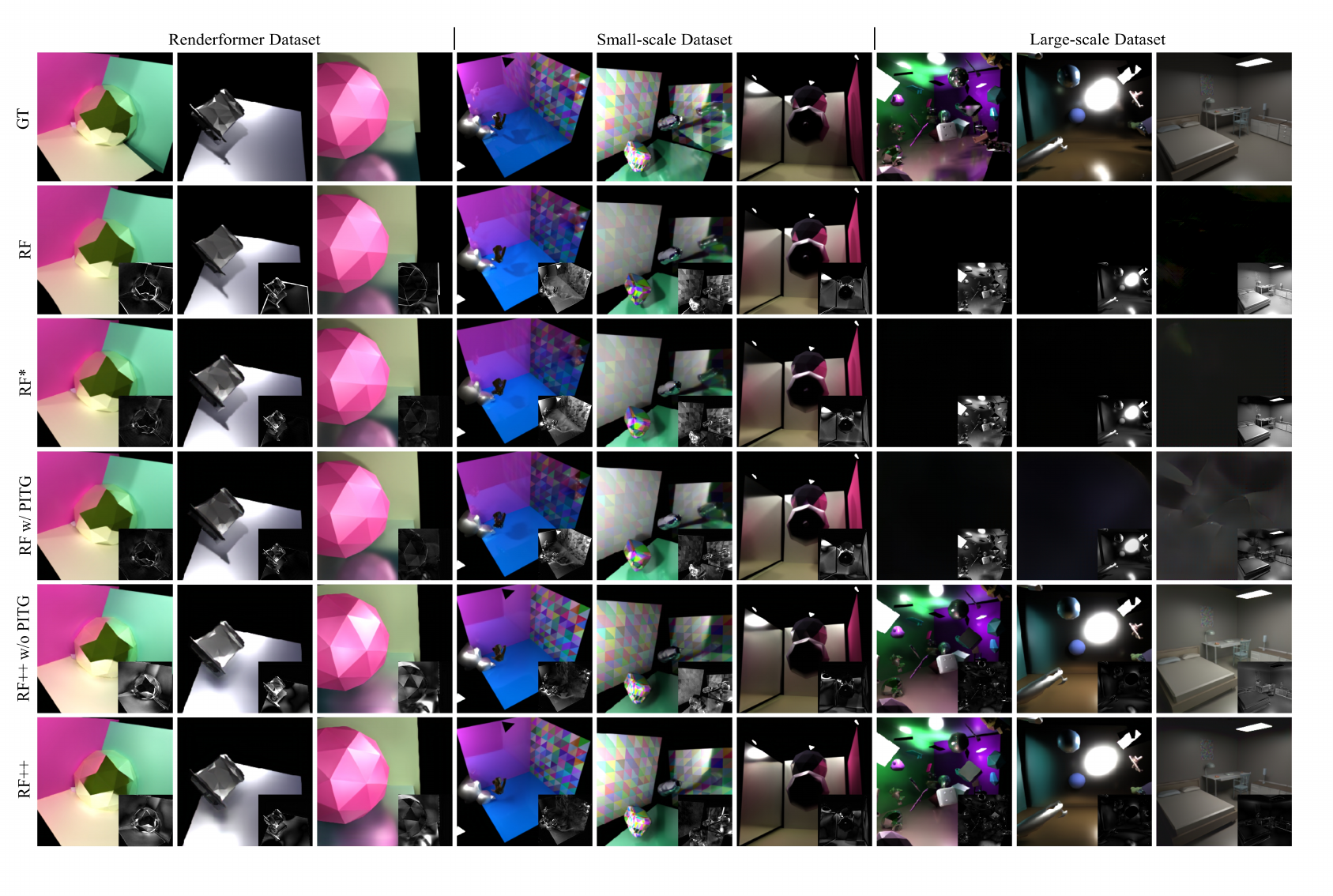}
  \caption{Qualitative comparisons on the test splits of the RenderFormer dataset and our small- and large-scale datasets; all scenes are unseen during training. Rows show the ground truth (GT), RenderFormer (RF), fine-tuned RenderFormer (RF*), RenderFormer with PITG, RenderFormer++ without PITG, and the full RenderFormer++. Insets show corresponding local error visualizations. The results demonstrate that PITG improves rendering fidelity, while HOCT maintains comparable quality on small-scale scenes and enables RenderFormer++ to handle substantially larger scenes. On large-scale scenes, both RF and RF* exhibit severe prediction degradation.}
  \Description{Qualitative comparisons of RenderFormer variants and RenderFormer++ on the test splits of the RenderFormer dataset and our
small- and large-scale datasets}
  \label{fig:qualitative_results}
\end{figure*}

\section{Experiments}

\subsection{Experimental Setup}

\paragraph{Dataset.}


We construct a large-scale synthetic dataset of 500K scenes following the RenderFormer data generation pipeline~\cite{zeng2025renderformer}. Each scene randomly samples object instances, materials, lighting, and camera viewpoints, and contains 5--70 objects with approximately 200--1200 triangles each. Rendering every scene from four viewpoints yields 2M training images.

These randomized configurations provide diverse spatial layouts and lighting conditions. Reference images are rendered in Blender at $256 \times 256$ resolution using adaptive sampling with 4096 samples per pixel (spp), followed by denoising. Surface reflectance is modeled by a microfacet BRDF with a GGX normal distribution~\cite{walterMicrofacet}, using diffuse albedo, specular albedo, and roughness.


To evaluate scalability, we evenly split the dataset into \emph{small-scale} and \emph{large-scale} subsets of 250K scenes each. The former contains approximately 3K--10K triangles per scene, while the latter contains 10K--80K.



For evaluation, we define \emph{train} and \emph{test} splits, each containing 1{,}000 small-scale and 1{,}000 large-scale scenes with four views per scene. The train split contains newly rendered views of scene
instances used during training, with camera viewpoints
disjoint from those in the training set, whereas the test
split contains entirely unseen scene instances.

\paragraph{Training Details.}

We train the full model for 500k iterations using AdamW
with cosine learning-rate decay from $1.0 \times 10^{-4}$.
Training takes 7 days on 8 NVIDIA A100 80GB GPUs
with a batch size of 32, using Flash Attention
\cite{dao2023flashattention2} and bfloat16 precision.

\subsection{Main Results}

We report both reconstruction metrics (L1 and MAPE) and perceptual metrics (LPIPS and SSIM). These metrics are evaluated on HDR images spanning a wide dynamic range.

\begin{table*}[t]
\centering
\small
\resizebox{\textwidth}{!}{%
\setlength{\tabcolsep}{8pt}
\begin{tabular}{llcccccccc}
\toprule
Dataset & Method & \multicolumn{4}{c}{Train split} & \multicolumn{4}{c}{Test split} \\
\cmidrule(lr){3-6} \cmidrule(lr){7-10}
& & L1$\downarrow$ & MAPE$\downarrow$ & LPIPS$\downarrow$ & SSIM$\uparrow$
  & L1$\downarrow$ & MAPE$\downarrow$ & LPIPS$\downarrow$ & SSIM$\uparrow$ \\
\midrule
Small-scale dataset 
      & RenderFormer       & 0.0895 & 0.1345 & 0.0886 & 0.8394
                 & 0.0905 & 0.1383 & 0.0907 & 0.8332 \\
      & RenderFormer*       & \textbf{0.0746} & \textbf{0.1140} & 0.0820 & 0.8836
                 & \textbf{0.0776} & \textbf{0.1119} & 0.0807 & 0.8822 \\
      & RenderFormer++ & 0.0886	 & 0.1265 & \textbf{0.0590} & \textbf{0.9542}
                 & 0.0861 & 0.1270 & \textbf{0.0601} & \textbf{0.9479} \\
\midrule
Large-scale dataset 
      & RenderFormer       & 0.5449	 & 0.8551 & 0.4531	 & 0.3575
                 & 0.5402 & 0.9020 & 0.4470 & 0.3558 \\
    & RenderFormer*       & 0.2765 & 0.7694 & 0.4383 & 0.3718
                 & 0.2984 & 0.8073 & 0.4322 & 0.3703 \\
    & RenderFormer++ & \textbf{0.0860} & \textbf{0.3298} & \textbf{0.1769} & \textbf{0.8305}
                 & \textbf{0.0899} & \textbf{0.3294} & \textbf{0.1766} & \textbf{0.7820} \\         
\bottomrule
\end{tabular}
}
\caption{Quantitative evaluated on large and small scene subsets. L1 and MAPE are computed in linear HDR space, while LPIPS and SSIM are computed in the perceptual tone-mapped space. Note: The asterisk (*) indicates the fine-tuned version.}
\label{tab:large_small_quality}
\end{table*}


\begin{table}[t]
\centering
\small
\setlength{\tabcolsep}{1.7mm}
\resizebox{\columnwidth}{!}{%
\begin{tabular}{@{}lcccc@{}}
\toprule
Method & L1$\downarrow$ & MAPE$\downarrow$ & LPIPS$\downarrow$ & SSIM$\uparrow$ \\ \midrule
\multicolumn{5}{@{}l}{\textit{RenderFormer Dataset}} \\
RenderFormer* & 0.0761 & 0.1129 & 0.0814 & \textbf{0.8829} \\
RenderFormer w/ PITG & \textbf{0.0706} & \textbf{0.1071} & \textbf{0.0806} & 0.8775 \\ 
RenderFormer w/ HOCT & 0.0868 & 0.1246 & 0.0895 & 0.8552 \\ 
RenderFormer++ ($K=3$)& 0.0793 & 0.1208 & 0.0843 & 0.8600  \\ \midrule
\multicolumn{5}{@{}l}{\textit{Small- and Large-scale Datasets}} \\
RenderFormer++ w/o PITG & 0.0899 & 0.2303 & 0.1203 & 0.8743 \\
RenderFormer++ w/o TCL & 0.0883 & 0.2290 & 0.1183 & 0.8757 \\
RenderFormer++ w/o GGD & 0.1198 & 0.2451 & 0.1276 & 0.8088 \\ \cmidrule(lr){1-5}
RenderFormer++ ($K=1$) & \textbf{0.0875} & 0.2289 & 0.1196 & 0.8702 \\
RenderFormer++ ($K=3$) & 0.0876 & \textbf{0.2281} & \textbf{0.1182} & \textbf{0.8758} \\
RenderFormer++ ($K=5$) & 0.0877 & 0.2283 & 0.1183 & 0.8758 \\
\bottomrule
\end{tabular}
}
\caption{Ablation study on PITG, HOCT, and GGD. Unless otherwise specified, RenderFormer++ uses $K=3$ as its configuration. The asterisk (*) indicates the fine-tuned version.}
\label{Ablation}
\end{table}

Table~\ref{tab:large_small_quality} compares RenderFormer++ with RenderFormer, which is additionally fine-tuned on our small-scale dataset for fairness.
RenderFormer++ achieves stronger perceptual quality on small-scale scenes and consistent performance across the train and test splits. Unlike RenderFormer, whose triangle-level tokenization precludes large-scale training, HOCT enables RenderFormer++ to train on large-scale scenes with improved quality and scalability.

Figure~\ref{fig:qualitative_results} compares RenderFormer++ with RenderFormer variants on unseen test splits from the RenderFormer dataset and our small- and large-scale datasets; further results and details are provided in the ancillary supplementary material.

\subsection{Efficiency Analysis}
Under a controlled setting with 50 objects (25{,}600 triangles) per scene, RenderFormer++ achieves $3.60\times$ and $3.86\times$ speedups in inference and per-step training, respectively, while reducing peak training memory by 55.6\% and maintaining comparable inference memory. Detailed efficiency profiling is provided in the ancillary supplementary material.

\subsection{Ablation Study}

We evaluate the ablation variants on the test splits of the RenderFormer dataset and our small- and large-scale datasets. For experiments on the RenderFormer dataset, G-buffer inputs are disabled for all variants to follow the original RenderFormer protocol.

\textit{Effect of Physics-Informed Transport Guidance.}  To evaluate PITG under both triangle-level and object-level tokenization settings, we conduct experiments on the RenderFormer dataset and Small- and Large-scale datasets. Under the triangle-level setting, we integrate PITG into RenderFormer and train it on the RenderFormer dataset. Under the object-level setting, we compare RenderFormer++ with and without PITG on our small- and large-scale datasets. As shown in Table~\ref{Ablation}, PITG consistently improves L1, MAPE, and LPIPS under both settings. SSIM does not exhibit a consistent gain because it is not directly optimized by our training objective. These results demonstrate the benefit of incorporating rendering-equation-inspired inductive biases into light transport modeling. Further qualitative comparisons are provided in the ancillary supplementary material.

\textit{Effect of Transport Consistency Loss.} As shown in Table~\ref{Ablation}, removing TCL degrades reconstruction performance, showing it benefits iterative transport refinement during training. As a training-only component of PITG, TCL provides modest improvement without additional inference overhead.

\textit{{Effect of Transport Iterations $K$.}}
The transport iteration number $K$ controls the depth of recursive feature-space transport propagation. As reported in Table~\ref{Ablation}, we compare reconstruction performance under $K=1$, $K=3$, and $K=5$. Performance varies only slightly across different values of $K$. $K=3$ achieves the best MAPE and LPIPS and ties for the best SSIM, while $K=1$ achieves the best L1. Since additional iterations increase computational cost and $K=5$ provides no further overall improvement, we choose $K=3$ as a practical trade-off between quality and efficiency.

\textit{{Effect of Geometry-Guided Decoder.}}
As reported in Table~\ref{Ablation}, incorporating G-buffer geometric attributes into the decoder consistently improves reconstruction quality. These geometry-guided features provide additional structural cues that help the decoder better recover fine-grained illumination details and surface appearance. These attributes can be efficiently generated using a standard rasterization pass, incurring negligible overhead. Notably, RenderFormer++ maintains a parameter count comparable to RenderFormer while supporting substantially larger scenes.



\textit{Effect of Hierarchical Object-Centric Tokenization.} 
To isolate HOCT, we compare RenderFormer with and without HOCT under the same protocol on the RenderFormer dataset. As shown in Table~\ref{Ablation}, HOCT introduces measurable degradation across the reconstruction and perceptual metrics. Adding PITG partially recovers this loss. In return, HOCT substantially improves runtime and memory scalability and enables large-scale training, demonstrating an accuracy--efficiency trade-off.

\section{Conclusion}
In this work, we propose \textbf{RenderFormer++}, a scalable and physics-informed feed-forward neural rendering framework for global illumination in triangle-mesh scenes. By introducing \textbf{Physics-Informed Transport Guidance}, our method incorporates rendering-equation-inspired inductive biases into Transformer-based light transport modeling. In addition, \textbf{Hierarchical Object-Centric Tokenization} aggregates triangle-level features into compact object-level representations, reducing the computational cost of global transport modeling for complex scenes. Experimental results and ablation studies demonstrate the effectiveness and scalability of the proposed framework.

\textit{Limitations and Future Work.}
The current dataset does not include textured objects, and material properties are shared across triangles within each object. Preliminary experiments on simple textured scenes, provided in the ancillary supplementary material, further suggest that the encoding scheme is compatible with richer appearance variations. In future work, we plan to extend the feature encoding module to support more diverse material properties and textured objects while maintaining scalability.

\bibliography{main}

\end{document}